\documentclass{article}

\usepackage{authblk}                                                           
\usepackage{hyperref}                                                          
\usepackage{graphicx}                                                          
\usepackage{subcaption}
\usepackage[font=footnotesize]{caption}
\usepackage{float}
\usepackage{placeins}
\usepackage{ifthen}
\usepackage[numbers]{natbib}  													
\usepackage{microtype}
\usepackage{physics}

\usepackage{xcolor}
\usepackage{soul}

\usepackage[normalem]{ulem}
\usepackage{pdfpages}

\newcommand{\figref}[2]{
	\ifthenelse{\equal{#1}{}}
	{
		\ifthenelse{\equal{#2}{}}
		{} 
		{(\subref{#2})}
	}
	{
		\ifthenelse{\equal{#2}{}}
		{\ref{#1}}       
		{\ref{#1}\,(\subref{#2})} 
	}%
	\unskip
}

\title{\ Tuning the Electronic Structure of Graphene by Controlling Spatial Confinement }
\author[1,2]{Mohammadamir Bazrafshan \thanks{Corresponding author: \href{mailto:mohammadamirbazrafshan@gmail.com}{mohammadamirbazrafshan@gmail.com}}}
\author[1,2,3]{Thomas D. Kühne}

\affil[1]{\small {Center for Advanced Systems Understanding (CASUS), Untermarkt 20, D-02826 Görlitz, Germany}}

\affil[2]{\small {Helmholtz Zentrum Dresden-Rossendorf, Bautzner Landstraße 400, D-01328 Dresden, Germany}}

\affil[3]{\small {Institute of Artificial Intelligence, Technische Universität Dresden, Nöthnitzer Straße 46, D-01187 Dresden, Germany}}

\date{\today}

\begin{document}
	\maketitle
	
	\begin{abstract}
		The electronic properties of a material depend on the spatial freedom of the electron wavefunction. A well-known example is graphite, which is a conventional gapless semiconductor, while a single layer of it, graphene, exhibits extremely high electronic conductivity. Nevertheless, graphene ribbons can have different physical properties, such as a tunable band gap, ranging from gapless to a large band gap semiconductor. The purpose of this study is to investigate the electronic structure of few-layer graphene composed of a layer of graphene nanoribbons and graphene sheet(s), where quasi-one-dimensional nanoribbons can interact with a two-dimensional sheet of graphite. Using the tight-binding model for graphite, we show how different configurations of such heterostructures can affect the electronic structure, which is different from that of their components. Our results show that systems composed of semiconducting AGNRs can not be seen as two separate systems.  Namely, a local gap of $\sim$0.6 eV at the Dirac point for dispersive bands can be opened in a bilayer configuration composed of a layer of gapless armchair nanoribbon stacked on graphene. We demonstrate that the band steepness in these structures can be tuned, highlighting their potential for electronic applications.
	\end{abstract}
	
	
	\section{\label{Intro}Introduction}
	
	Quantum mechanics deepens the understanding and prediction of material properties. The pure classical particle interpretation of the electron fails to explain several phenomena observed in materials, whereas the wave-based quantum interpretation can, and it also predicts novel properties such as the Berry phase~\cite{Berry1984}. Furthermore, the spatial confinement of the electron wavefunction can change the properties of a material. For example, graphite is composed of single layers of carbon atoms stacked primarily in the Bernal (ABA) configuration. Although this natural form of carbon has existed for billions of years, the remarkable physical properties of its individual layers, known as graphene, remained unexplored experimentally until their isolation in 2004~\cite{Novoselov2004,Geim2009}. This breakthrough revealed unique quantum phenomena such as massless Dirac fermions and the anomalous half-integer quantum Hall effect, sparking widespread interest in two-dimensional (2D) materials~\cite{CastroNeto2009, Dean2012, Geim2013}. Moreover, as the range of available 2D materials expanded, the mixing of materials with different properties into heterostructures\cite{Jia2022,Qin2023,daqiqshirazi2024,Lv2024} became an active field as the physical understanding of the materials progressed.   
	
	Although both graphite and graphene are zero-gap semiconductors (semimetals), their electronic conductivities are substantially different due to their electronic structure~\cite{Slonczewski1958,Reich2002}. 
    Electrons in graphite are not confined, whereas in graphene, they are confined in 2D. Graphite has parabolic bands due to interlayer coupling, while graphene has linear bands, that touch at the Dirac (K) point close to the Fermi energy ($E_F$), leading to an extraordinary carrier mobility and ballistic transport~\cite{Rurali2017}. 
	
	The tight-binding (TB) approximation provides an efficient way to capture quantum mechanical effects, thanks to its small and localized basis set. This makes it especially useful for gaining insight or performing initial screening in large systems, where more accurate methods become computationally demanding~\cite{Slater1954}. 

	Although graphene exhibits linear bands originating from its lattice symmetry and $\pi$-electron network, bilayer graphene has parabolic bands (near $E_F$), whereas its trilayer shows a combination of linear and parabolic bands.~\cite{Grneis2008}. However, the electronic properties of graphene nanoribbons (GNRs), which are quasi-1D systems, can vary depending on their width and edge geometry. For example, armchair GNRs (AGNRs) can behave as a semiconductor or a semimetal depending on their width, while zigzag GNRs (ZGNRs) are always semimetallic due to their highly localized edge states, as predicted using the simple TB method~\cite{CastroNeto2009,Son2006}. Moreover, graphene sheet is almost transparent to visible light~\cite{Mak2008}, but GNRs are not necessarily transparent~\cite{Denk2014}.
    As the electronic structure of graphene and its nanoribbons is different~\cite{Houtsma2021}, and motivating the effect of geometries, which comes from the confinement effect, we combined quasi-1D GNRs in Bernal-stacked bilayer and trilayer configurations, to create a graphene heterostructure. 
    We only studied ABA stacking, and for the sake of brevity, we did not mention it specifically hereafter. However, ABC and ABA stacking are both stable and have a small total energy difference~\cite{GuerreroAvils2022,Aoki2007}, but ABA stacking is observed more frequently~\cite{Hattendorf2013,CastroNeto2009}.

    Using the TB method, we found that an array of GNRs can be used to induce a variety of modifications in the electronic structure. Depending on the GNRs considered, these heterostructures can behave in rather different ways, ranging from opening a local gap to controlling the curvature of the energy band close to $E_F$. The interlayer interaction in graphite is several times weaker than the intralayer interactions, meaning that one can predict that narrow semiconducting GNRs should not have a strong impact, while semimetal ones should. While in some configurations this seems to work near $E_F$, but there are some other configurations that doesn't consistent with this picture. However, the details are very interesting; a bilayer composed of a semimetal AGNR and graphene can have a local gap (at K) of $\sim$0.6 eV for dispersive band. The case for a trilayer with an array of semimetal AGNRs as the middle layer leads to the disappearance of parabolic bands of the perfect trilayer, yet the linear-like bands remain. Moreover, if the layer switches with one of its neighboring layers and, for semiconducting AGNRs, only the parabolic bands of the trilayer remain, or one can say that it behaves as a bilayer. In contrast, for semimetal AGNRs, the band steepness becomes larger than that of the parabolic bands of the perfect trilayer, which can be beneficial for conductivity. This behavior happens for the other trilayer configuration by widening the semiconducting AGNRs. These results shows the potential of these novel systems. 
	
	This manuscript is organized as follows: In Section~\ref{Model}, we discuss the model and the theoretical method used to simulate the systems. The numerical results, together with the discussion, are presented in Section~\ref{Results} and we conclude our findings in the last section.
    
	\section{\label{Model}Model and Method} 
    
\begin{figure}[tp]
		\centering
		\begin{subfigure}[t]{0.62\linewidth}
			\caption{}
			\includegraphics[height=4cm]{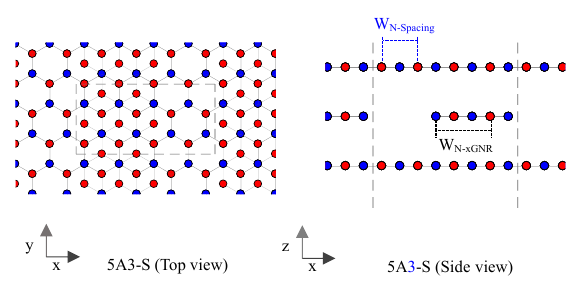}
			\label{1a}
		\end{subfigure}
		\hfill
		\begin{subfigure}[t]{0.30\linewidth}
			\caption{}
			\includegraphics[height=4cm,trim={0.1cm 0 0 0},clip]{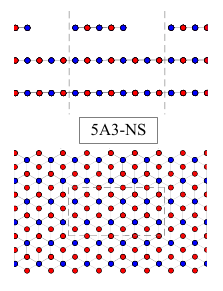}
			\label{1b}
		\end{subfigure}
		\caption{(a) The top view (left) and side view (right) of the 5A3-S atomic structure with our naming convention. (b) The side view (top) and top view (bottom) of the 5A3-NS model, where the AGNR array is not sandwiched. The unit cell borders are shown by dashed lines, and the two sublattices are shown by red and blue colors. }
		\label{fig1}
	\end{figure}

	The atomic structure of two trilayer heterostructures with sandwiched and non-sandwiched GNR arrays is shown in Figure~\figref{fig1}{1a}and~\figref{}{1b}, together with our naming convention. The naming convention for a trilayer heterostructure consists of four parts. The first part, which is a number, indicates the width of the GNR ($\mathrm{W_{N-xGNR}}$), counting the number of atoms in the width. The second part is a character that indicates the type of GNR: \textbf{A} for armchair and \textbf{Z} for zigzag edge nanoribbons. The third part is a number that determines the spacing between the GNRs ($\mathrm{W_{N-Spacing}}$), similar to the GNR width. Finally, after a dash, the arrangement of the GNR array is indicated by a character. If the array is sandwiched between two graphene layers, \textbf{S} is shown; otherwise, \textbf{NS} is shown, see Figure~\figref{fig1}{}. The last part of the bilayer configuration is meaningless and can be neglected or omitted, as in this work. 
    
	The six-parameter Slonczewski–Weiss–McClure (SWMcC) TB model for graphite is used~\cite{Slonczewski1958,McClure1957}, as it can effectively justify experimental observations~\cite{Lui2011,Geim2013}.
    The electronic properties of the heterostructures are studied by employing the $p_z$ orbital real-space TB model with the Hamiltonian that reads:
    \begin{equation*}
    	  H = -\sum_{i \neq j} t_{ij} \left( |i\rangle \langle j| + |j\rangle \langle i| \right) + \sum_{i \in \alpha \cup \beta } \Delta_i |i\rangle \langle i| 
    \end{equation*}
    where $i$ and $j$ denote atomic sites and $t_{ij}$ are the hopping integrals. The on-site energies $\Delta_i$ account for sublattice asymmetry, which arises in few-layer graphene due to the crystal field. These on-site energies are nonzero only on specific sites, defined by the sets $\alpha$ and $\beta$, which correspond to the two sublattices in the lattice. An alternative convention assigns $+\Delta/2$ to one sublattice and $-\Delta/2$ to the other; both conventions are algebraically equivalent up to a uniform energy shift and yield the same physical results. Within this model, the hopping parameters $t_{ij}$ are given by $\gamma_0, \gamma_1, \gamma_2, \gamma_3, \gamma_4,$ and $\gamma_5$, corresponding to intra- and interlayer hopping processes in Bernal-stacked graphite or multilayer graphene. The on-site term $\Delta_i$, represents the difference in potential between nonequivalent sublattices ,    and the TB parameters are indicated in the Figure \figref{figs1}{} in the Supplementary Materials (SM). The details of the SWMcC model can also be found elsewhere~\cite{Charlier1991,Slonczewski1958,McClure1957,Dresselhaus1981}.
	
	\section{\label{Results}Results and Discussion}
    
	As mentioned, bilayer graphene has two parabolic bands because of interlayer interactions, while trilayer graphene has linear bands in addition to the parabolic bands. However, the strongest interlayer interaction (with hopping term $\gamma_1=0.39$ eV) is much weaker than the intralayer interaction (with $\gamma_0=3.16$ eV). Nevertheless, such weak interactions can substantially alter the low-energy electronic structure near the K point. In the following, we study and discuss the numerical results for the electronic structure of bilayer and trilayer heterostructures. To be more realistic, the minimum spacing between AGNRs is considered to be 2, and 4 for ZGNRs in the layer where GNRs are arrayed. 
    
    It is known that the band gap of an AGNR can be classified into 3 groups, namely, $\mathrm{3p+2<3p<3p+1}$ with p being an integer~\cite{Son2006,Bazrafshan2026}. The 3p+2 group is the semimetal or the gapless group. This classification is essential for systematically organizing the studied structures based on the electronic character of their constituent AGNRs.
	
	The simple TB method is a semi-empirical model to justify the electronic properties of the systems in a limited energy range, hereafter we only discuss the energies close to $E_F$. We should note that single- and few-layer graphene are highly sensitive to structural symmetries, and breaking that symmetry can significantly modify the electronic structure, e.g., by opening a band gap at K or altering the Dirac cones in energy or momentum space, as well as through other modifications \cite{CastroNeto2009,Zhou2007,McCann2013,Kindermann2011,Cheng2010}. In graphene and GNRs within a limited energy range near $E_F$ (or the band-gap edges), electron–hole symmetry exists, as predicted by the simple TB description. For the sake of brevity, we therefore refer only to the valence bands hereafter.
	
	\subsection{Trilayer with a sandwiched GNR array} 
	       
\begin{figure}[ht]
		\centering
		\begin{subfigure}[t]{0.36\linewidth}
			\caption{}
			\includegraphics[height=4.8cm,trim={5cm 7.7cm 5.9cm 7.7cm},clip]{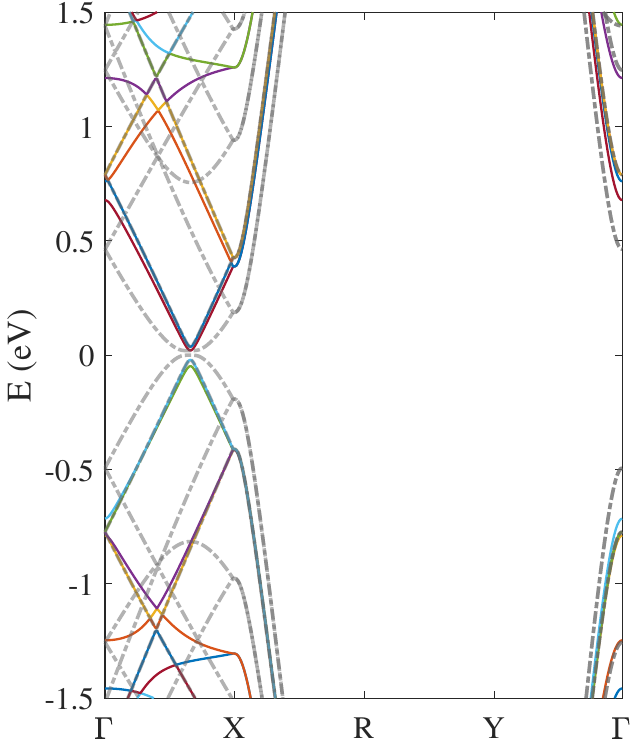}
			\label{2a}
		\end{subfigure}
		\hfill
		\begin{subfigure}[t]{0.31\linewidth}
			\caption{}
			\includegraphics[height=4.8cm,trim={6.6cm 7.7cm 5.9cm 7.7cm},clip]{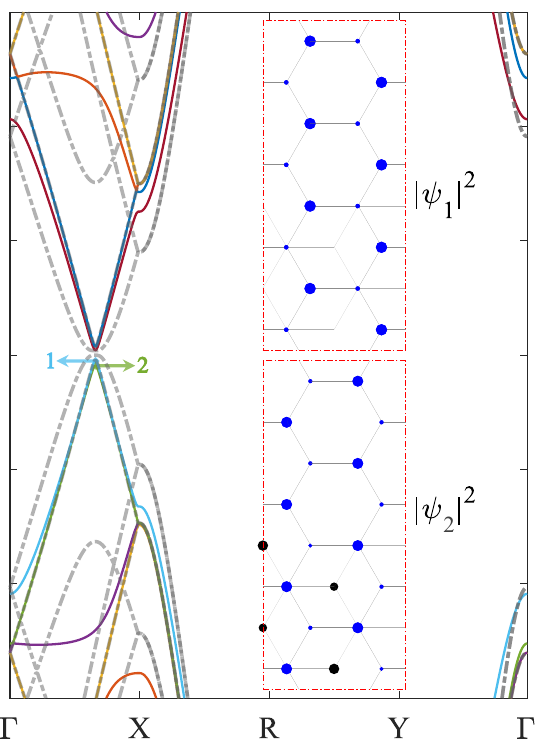}
			\label{2b}
		\end{subfigure}
		\hfill
		\begin{subfigure}[t]{0.31\linewidth}
			\caption{}
			\includegraphics[height=4.8cm,trim={6.6cm 7.7cm 5.9cm 7.7cm},clip]{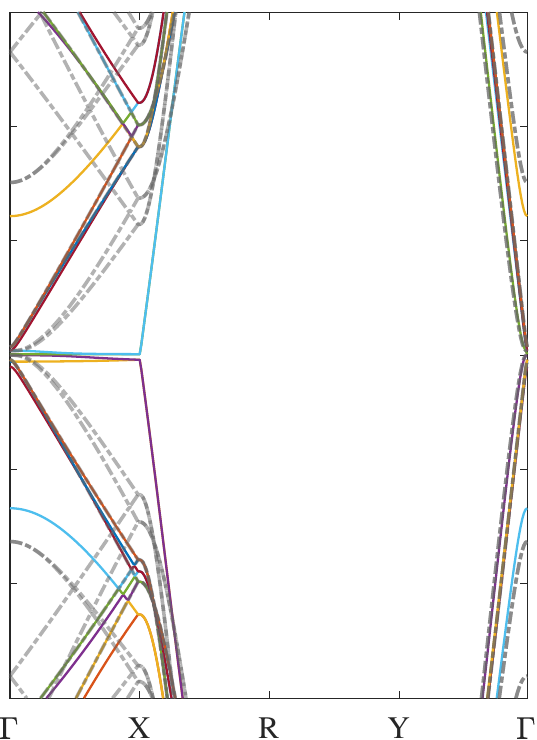}
			\label{2c}
		\end{subfigure}
		\caption{The electronic band structure of (a) 3A4-S, (b) 4A4-S and (c) 5A4-S heterostructures along selected Brillouin zone points. black filled circles for the middle layer, and blue is for other layers. The band structure of the corresponding perfect systems are shown by dashed gray lines for reference.}
		\label{fig2}
	\end{figure}

	We first study the width dependence of three AGNRs, covering all AGNR groups, with a constant spacing. In Figure~\figref{fig2}{}, the electronic band structure of a trilayer with a sandwiched array of 3-AGNRs (3A4-S) \figref{}{2a}, 4-AGNRs (4A4-S)\figref{}{2b} and 5-AGNRs (5A4-S)\figref{}{2c}, all with 4 spacings, is shown together with the corresponding perfect cases (dashed gray lines). An AGNR with a width of 3 or 4 atoms is a semiconductor, meaning that it does not have available electronic states near $E_F$. For states close to  $E_F$ in graphene, this implies that they cannot effectively interact with such AGNRs. In this respect, sandwiching a gapped material between graphene sheets is similar to studying individual graphene sheets. The band structure shows very close but not degenerate bands, which does not completely support the above-mentioned insight. 
    
    The wavefunctions of two selected states from the first (state 1) and second (state 2) bands at K further highlight this contrast. One state shows no contribution from the middle layer, whereas the other state does, as can be seen in the inset of the Figure~\figref{fig2}{2b}. Contributions from the middle layer are shown in black, and all other contributions in blue, with this color code used hereafter. Therefore, the two outer layers can interact through the middle layer even though it is a semiconducting AGNR, showing that treating them as a system composed of individual, non-interacting layers is not correct. However, because the bands are very close to each other and are similar to those of graphene, one may conclude that the interaction is not strong, but the absence of flat bands, which would be expected if the GNRs were not effectively coupled, further supports the view that the system behaves as a whole. Moreover, the spacing between AGNRs does not observed to have a strong impact on the electronic structure, see Figure~\figref{figs2}{} in the SM. 
	
	For the 5A4-S system, which contains a semimetallic AGNR, the electronic band structure shows flat bands with dispersive bands at different $\mathbf{k}$-paths (see Figure\figref{fig2}{2c}). The transverse electronic modes in an AGNR are quantized (they cannot have continuous $\mathbf{k}$ values), appear as a flat band along $\Gamma-$X $\mathbf{k}$-path. However, along the periodic nanoribbon directions (X$-$R and Y$-\Gamma$),  system behave similar to the isolated 5-AGNR.
	
\begin{figure}[h]
		\centering
		\begin{subfigure}[t]{0.35\linewidth}
			\caption{}
			\includegraphics[height=4.8cm,trim={5cm 7.7cm 5.9cm 7.7cm},clip]{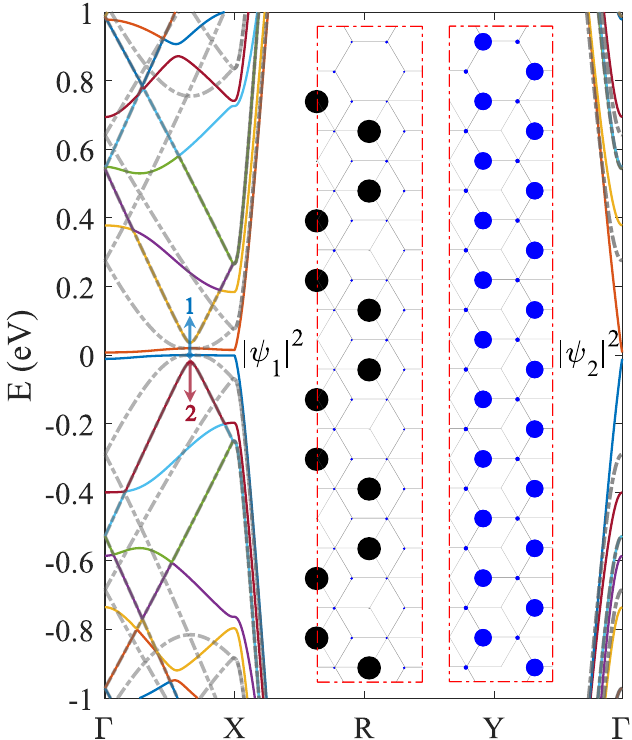}
			\label{3a}
		\end{subfigure}
		\hfill
		\begin{subfigure}[t]{0.31\linewidth}
			\caption{}
			\includegraphics[height=4.8cm,trim={6.6cm 7.7cm 5.9cm 7.7cm},clip]{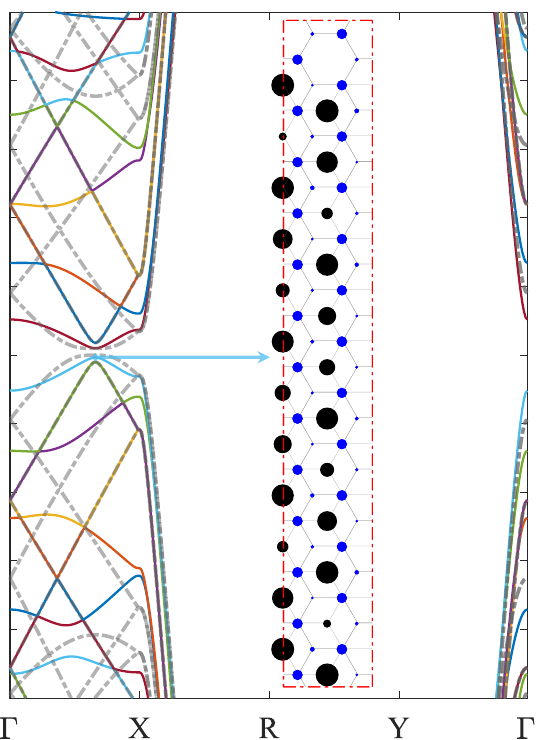}
			\label{3b}
		\end{subfigure}
		\begin{subfigure}[t]{0.31\linewidth}
			\caption{}
			\includegraphics[height=4.8cm,trim={6.6cm 7.7cm 5.9cm 7.7cm},clip]{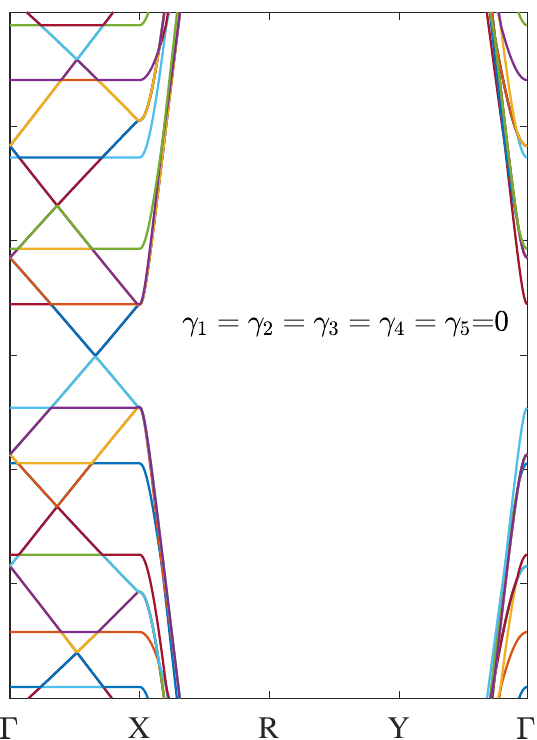}
			\label{3c}
		\end{subfigure}		
		\caption{The electronic band structure of wide (a) semimetal 20A2-S, and (b) semiconducting 24A2-S. The site-resolved probability amplitude for the marked states are shown in the inset. (c) The electronic band structure of 24A2-S without interlayer interactions.}
		\label{fig3}
	\end{figure}

	We should imply that in AGNRs, the electronic allowed states are not localized on the edges\cite{Kleftogiannis2013,Brey2006}, indicating that edge-induced effects are suppressed in wider ribbons. In Figures~\figref{fig3}{3a} and \figref{}{3b}, the electronic band structure is shown for systems containing gapless (semimetal) and semiconducting AGNRs, respectively. 
    For the semiconducting AGNRs, the bands become parabolic and tend to be more similar to the perfect systems as the width of the ribbon increases. We also studied $\abs{\psi}^2$ for two states marked on the band structure of the 20A2-S system and show that the state selected from the flat band is concentrated in the middle layer, while the dispersive one (both at K) is distributed across the neighboring layers, see the inset of Figure~\figref{fig3}{3a}.

	Furthermore, examining the band structure of the 24A2-S system, with wider semiconducting AGNRs, (Figure~\figref{fig3}{3b}) reveals that the coupling of the sandwiched semiconducting AGNR array becomes stronger, as by the modified dispersion of the first valence band. 
    The $\abs{\psi}^2$ of the first valence band at K, indicates that the middle layer has a larger portion of the wavefunction, but other layers also host a part of it. We also decouple the layers by considering only the in-plane (intra-layer) hopping term ($\gamma_0$) and setting other hopping terms to zero; see Figure~\figref{fig3}{3c}, where the linear bands are doubly degenerate, and the flat bands are due to the AGNR array (along $\Gamma-$X). This helps show what should be expected in the case of non-interacting layers, i.e., one should be able to see the band structure of the AGNR alongside that of graphene. From panels \figref{}{3b} and \figref{}{3c} of Figure~\figref{fig3}{}, it can be concluded that the interlayer couplings play a significant role in tuning the band dispersion for this system. In particular, the steepness of the first valence band reduces comparing to narrow AGNRs, compare Figure~\figref{fig2}{2a} (or Figure~\figref{fig2}{2b}) with Figure~\figref{fig3}{3b}. Therefore, while it seems that a non-interacting picture works reasonably for a limited energy range for narrow semiconducting AGNRs, it fails for wider ones.  
    
\begin{figure}[h]
		\centering
			\begin{subfigure}[t]{0.3\linewidth}
				\caption{}
				\includegraphics[height=4.8cm,trim={5cm 7.0cm 5.9cm 7.7cm},clip]{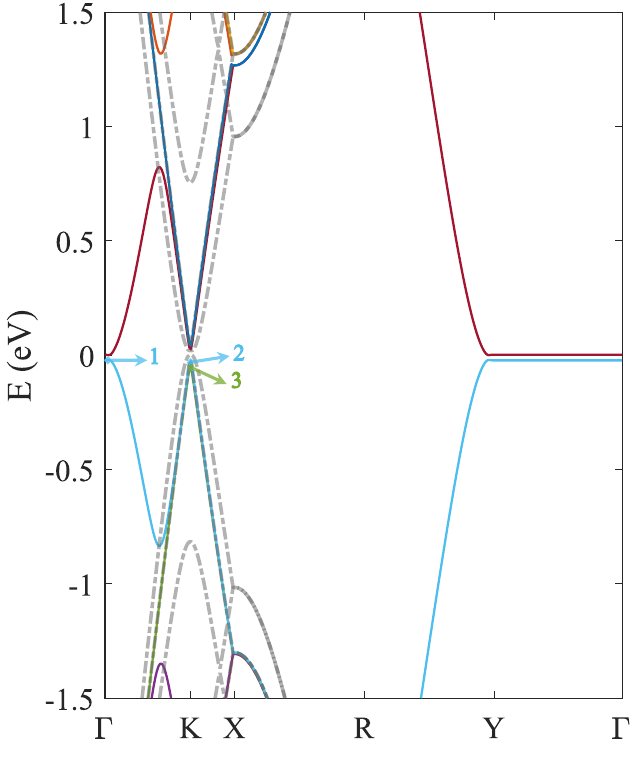}
				\label{4a}
			\end{subfigure}
			\hfill 
			\begin{subfigure}[t]{0.6\linewidth}
				\caption{}
				\includegraphics[height=4.8cm]{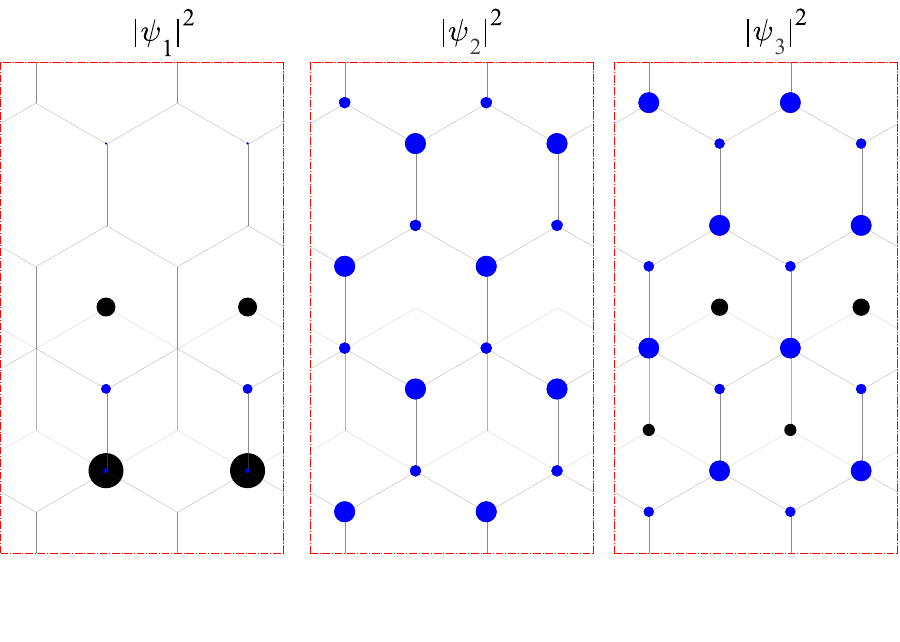} 
				\label{4b}
			\end{subfigure}
		\caption{The band structure of (a) 4Z4-S and (b) $\abs{\psi}^2$ for the selected states (marked on panel (a)).}
		\label{fig4}
	\end{figure}

	The ZGNRs are known for their highly localized edge states, where widening the GNR width weakens the edge effect, suggesting that an array of narrow ZGNRs should not have a strong impact on the electronic structure, limiting the coupling with the neighboring layers. The result for 4Z4-S is shown in Figure~\figref{fig4}{4a}, with two linear-like degenerate bands at {the K point}. Such a degeneracy suggests that there are two nearly isolated graphene sheets. To have a better insight, we selected three states for furthur investigation: state 1, chosen from the valence band at the $\Gamma$ point, resides primarily at the edge of the arrayed 4-ZGNR; state 2, chosen from the first valence band at the K point, has no contribution from the middle layer; and state 3, chosen from the second valence band at the K point, shows some contribution from the middle layer, see Figure~\figref{fig4}{4b}. The high dispersion of these bands indicates they arise from propagating states. Note that in the simple TB, edge effects exist for any width of a ZGNR with periodic boundary condition. 
    However, as the width of the nanoribbons increases, the coupling between the layers also increases, and the system tends to behave as a trilayer, see Figure~\figref{figs3}{} in the SM. 

    Taken together, it is shown that although narrow AGNRs in this configuration do not strongly modify the electronic structure close to $E_F$, the absence of flat bands and the $\abs{\psi^2}$ analysis show that the middle layer is coupled to the system, in which wider ribbons have different valence band steepness. Particularly, the case with semiconducting AGNRs is interesting because embedding a layer of arrayed AGNRs in a trilayer structure produces two highly dispersive valence bands without requiring spatial isolation, or what we can call it thick graphene. 
    In contrast, achieving similar dispersive bands in conventional few-layer graphene typically requires separating the layers, which may increases the vertical spacing and can introduce mechanical instabilities, making such free-standing layers highly challenging to realize. The presence of non-symmetry-related dispersive bands observed in S configuration heterostructures containing semimetal class of AGNRs or ZGNRs, reflects the existence of multiple delocalized propagation channels.

	\newlength{\figwidth}
	\setlength{\figwidth}{0.7\linewidth} 
	
	\subsection{Trilayer with a non-sandwiched GNR}
	
\begin{figure}[h]
		\centering
		\begin{subfigure}[t]{0.35\linewidth}
			\caption{}
			\includegraphics[height=4.8cm,trim={5cm 7.7cm 5.9cm 7.7cm},clip]{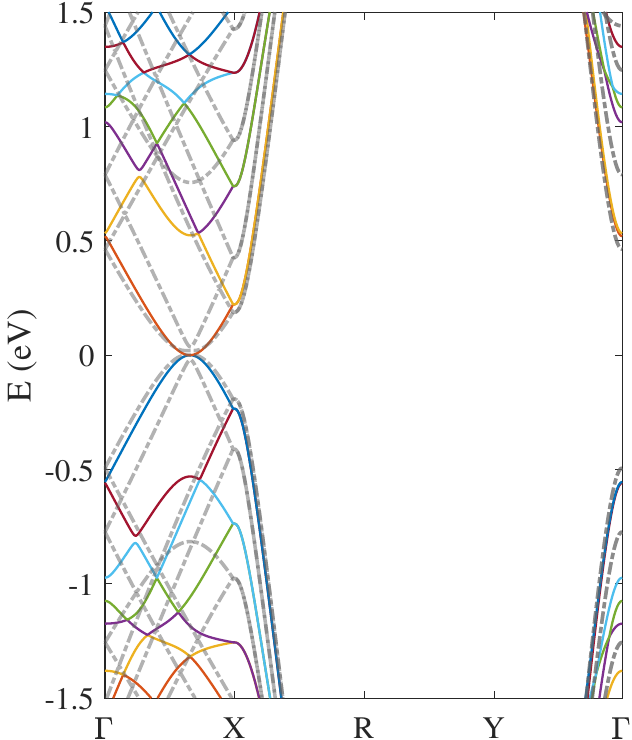}
			\label{5a}
		\end{subfigure}
		\hfill
		\begin{subfigure}[t]{0.31\linewidth}
			\caption{}
			\includegraphics[height=4.8cm,trim={6.6cm 7.7cm 5.9cm 7.7cm},clip]{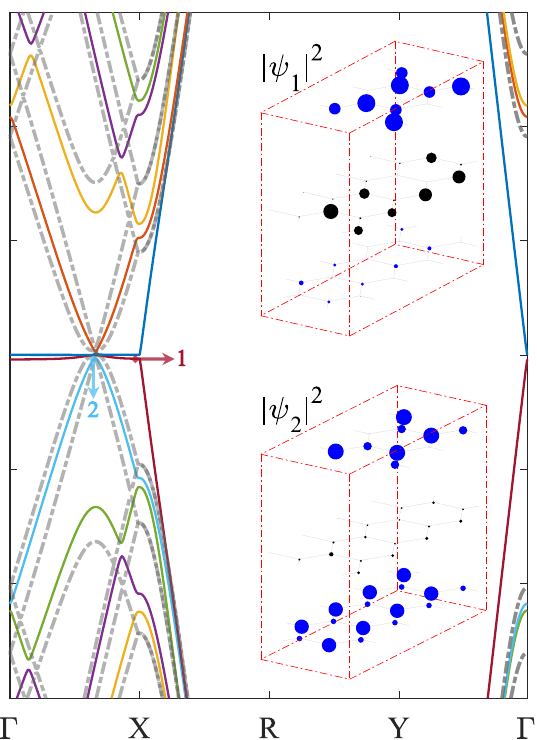}
			\label{5b}
		\end{subfigure}		
		\begin{subfigure}[t]{0.31\linewidth}
			\caption{}
			\includegraphics[height=4.8cm,trim={6.6cm 7.7cm 5.9cm 7.7cm},clip]{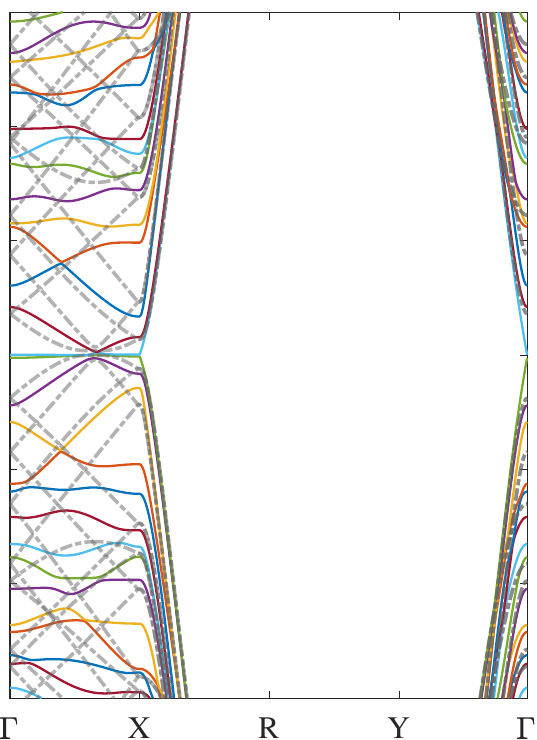}
			\label{5c}
		\end{subfigure}	
		\caption{The electronic band structure of (a) 4A3-NS, (b)5A3-NS (with $\abs{\psi}^2$ for the marked states), and (c) 23A3-NS systems.}
		\label{fig5}
	\end{figure}

	Having insight from sandwiching an array of GNRs between graphene layers brings to mind the following question: What would happen if the array is placed in the system instead of one of the outer-layers, i.e., in the NS configuration? We know that the electronic properties of graphene are governed by the out-of-plane $2p_z$ orbital. For the S configuration, the middle layer symmetrically interacts with neighboring layers, while for the NS configuration, interactions are different, meaning one graphene layer has a stronger coupling with the GNR array.  
	In Figure~\figref{fig5}{5a} and \figref{}{5b}, the electronic band structures for a semiconducting AGNR in 4A3-NS and a gapless one in 5A3-NS heterostructures are shown, respectively.  While the 4A3-NS system shows the electronic structure of perfect bilayer graphene with parabolic bands, there is no clear signature of an isolated semiconducting AGNR, indicating that the AGNR layer effectively interacts with other layers. The 5A3-NS system exhibits a different behavior, where a band between the parabolic and linear bands of the perfect system appears and remains present for the wider case (see Figure~\figref{fig5}{5c}).This shows that the change in band steepness is not affected by width. In this context, we studied $\abs{\psi}^2$ for two selected bands: $\abs{\psi_1}^2$ is from the flat band at X, and $\abs{\psi_2}^2$ is from the dispersive band at the band-touching point K, for the 5A3-NS system, as shown in the inset of Figure~\figref{fig5}{5b}. Analysis indicates that $\abs{\psi_1}^2$ originates mainly from the GNR array and its neighboring layer, while $\abs{\psi_2}^2$ is primarily localized on the outer layers, confirming that the electronic structure changes significantly when a gapless AGNR is present in the system. 
    
    We already seen that the band steepness changes when a layer of semiconducting AGNRs is in the S configuration, in which first valence band's steepness changes but the second one doesn't and stays similar to the one of the perfect trilayer (see Figure~\figref{fig3}{3b}). However, here one can see band tuning happens for the NS heterostructures when a layer of gapless AGNRs is present.

	A system containing ZGNRs shows predictable behavior in the electronic structure, at least close to $E_F$, see Figure~\figref{figs4}{s4a} and \figref{}{s4b} respectively for 4Z4-NS and the wider 10Z6-NS in the SM. The system is composed of a bilayer graphene with an array of ZGNRs, such that in the electronic band structure one can see bands similar to the parabolic bands of the perfect trilayer system (around K). This can also be understood from the probability amplitude of the marked states in the band structure (see the inset of Figure~\figref{figs4}{s4a}). However, for higher energies, it can be seen that the first and second valence/conduction bands show a gap at the same $\mathbf{k}$-vector (indicated by the double arrow), a property that can influence excitations in the system, which is beyond the scope of this work.

	\subsection{Bilayer GNR-Graphene}
	
\begin{figure}[h]
		\centering
		\begin{subfigure}[t]{0.35\linewidth}
			\caption{}
			\includegraphics[height=4.8cm,trim={5cm 7.7cm 5.9cm 7.7cm},clip]{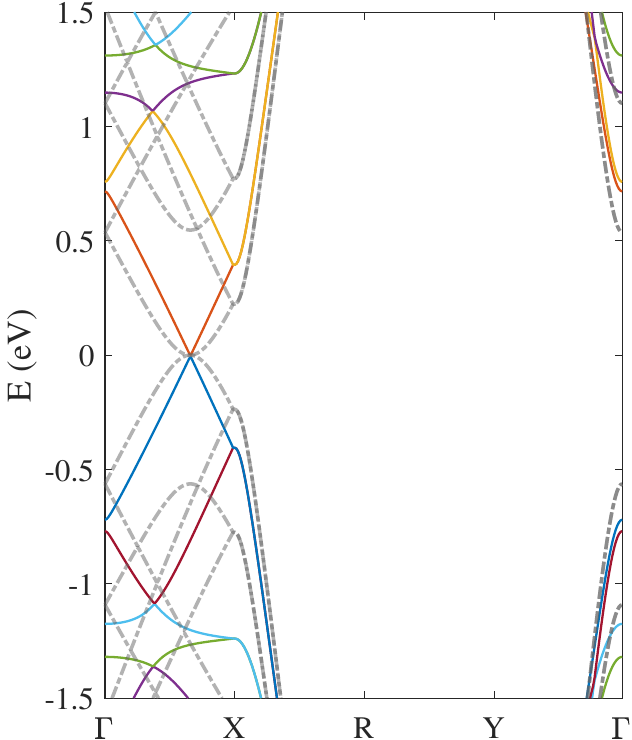}
			\label{6a}
		\end{subfigure}
		\hfill
		\begin{subfigure}[t]{0.31\linewidth}
			\caption{}
			\includegraphics[height=4.8cm,trim={6.6cm 7.7cm 5.9cm 7.7cm},clip]{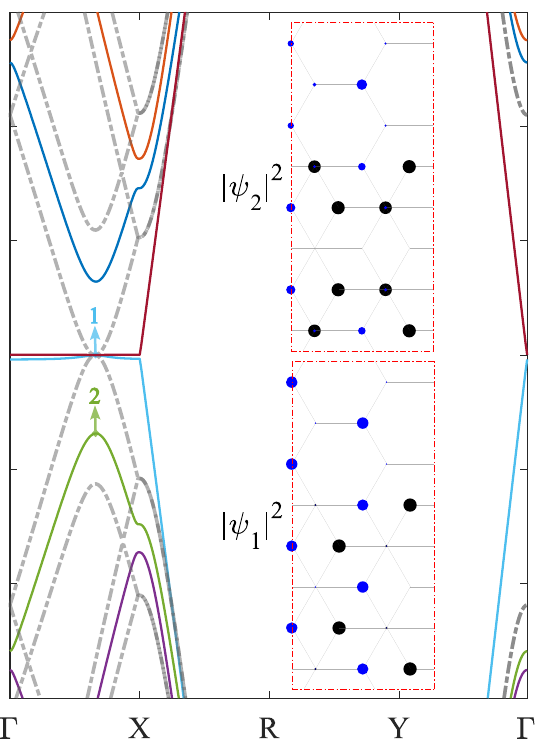}
			\label{6b}
		\end{subfigure}		
		\begin{subfigure}[t]{0.31\linewidth}
			\caption{}
			\includegraphics[height=4.8cm,trim={6.6cm 7.7cm 5.9cm 7.7cm},clip]{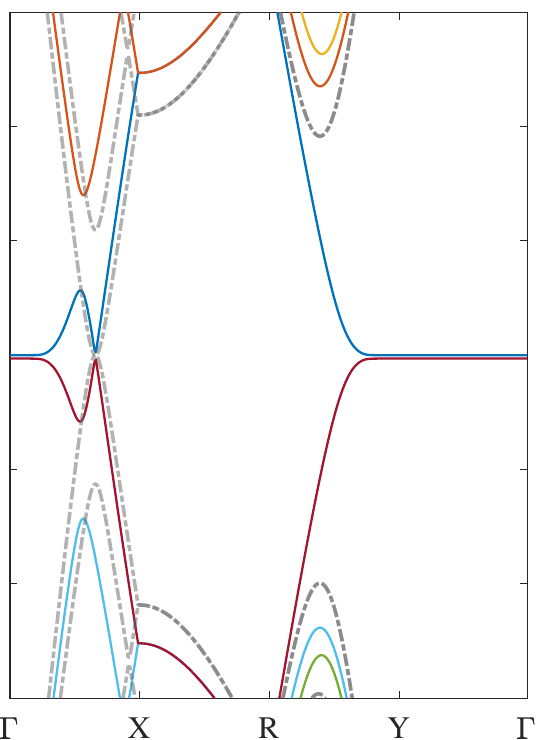}
			\label{6c}
		\end{subfigure}	
		\caption{The electronic band structure of (a) 4A3, (b) 5A3, and (c) 12Z4 bilayer heterostructures. In panel (b), two states and their corresponding spatial wavefunction distribution are shown. }
		\label{fig6}
	\end{figure}

	In this subsection, we study a heterostructure consisting only one array of GNRs stacked on graphene. From the point of view of individual non-interacting layers, we expect that placing a semiconducting material on graphene will not strongly modify the electronic structure,especially near $E_F$, as confirmed by the electronic band structure of the 4A3 system (see Figure~\figref{fig6}{6a}), where linear bands of graphene are present, i.e., the system is essentially a graphene monolayer. As in previous systems, the absence of flat bands at the band edges of the isolated AGNR indicates that the layers are electronically coupled. Instead, the observed energy bands reflect the allowed states of the heterostructure, which closely resemble the dispersive bands of graphene. For this system, we also applied an electric field and showed that a band gap develops, indicating that the system is not composed of two independent subsystems, while such a perturbation can produce a meaningful change in the electronic structure, see Figure \figref{figs5}{} in the SM.

	However, a gapless 5-AGNR induces a 0.6 eV local gap at K for the dispersive band, see Figure\figref{fig6}{6b}. 
    In the inset, $\abs{\psi_1}^2$ and $\abs{\psi_2}^2$ selected from the first and second valence bands, are shown, indicating that both states are spread over the system and are not localized within each of layers.  
    The effect of ZGNR arrays is studied in Figure\figref{fig6}{6c} for the 12Z4 system, where it is similar to previous systems in trilayer configurations. 
    
    As we have implied, while these simulations can effectively capture quantum mechanical effects, the influence of structural relaxation is not considered, which can modify the electronic properties of the systems~\cite{Khne2020}. For wider GNRs, the impact of relaxation is expected to be reduced, and the TB approach is expected to be sufficient to capture the main qualitative features discussed in this work. Nevertheless, edge originated geometrical modification is reduced upon hydrogenation, yet doesn't change the electronic structure strongly\cite{Son2006,Yu2015}. Interested readers can also see Figure~\figref{figs6}{} in the SM for further discussion on the local gap at K of the 5A11-NS trilayer heterostructure. 
	It is worth noting that we did not observe a notable impact of the spacing between GNRs on the electronic structure in any of the systems studied. 
	
	\section{\label{Conclusion}Conclusion}
	
	The electronic properties of bilayer and trilayer configurations of graphene heterostructures, with an array of armchair and zigzag GNRs, are studied using a six-parameter TB approximation for graphite. Our results show that when such an array of AGNRs is present in trilayer or bilayer configurations, depending on the width of the nanoribbon, two scenarios can be happened. If the AGNR is semiconducting and is placed as the middle layer in the trilayer configuration, two almost linear-like bands of the graphene layers are present in the band structure. However, for gapless AGNRs, there is only one linear band, showing that the whole system behavior becomes close to that of graphene, which, for the NS configuration, no linear-like band is present, yet a band that lies between the two characteristic bands of the perfect trilayer appears, indicating band steepness modification. This modification happens for S configuration with semiconducting class of AGNRs. Numerical simulations shows that the spacing between the GNRs doesn't has a strong effect on the electronic structure. 
	
	The bilayer systems are also studied. Our findings show that gapless AGNRs can open a local gap at K with a value of $\sim 0.6$ eV for dispersive bands. Interestingly, the system has a flat band, and the associated electron wavefunction is spread over the entire structure. Such an extended nature may enhance the chemical reactivity. Semiconducting AGNRs in this configuration doesn't change the electronic structure close to $E_F$, yet it is shown by applying a perturbation, a gap developed at K. 
	
	All studied heterostructures containing ZGNRs behave in a similar way within a limited range of energies close to $E_F$. 
    We show that the main electronic structure features, related to the class of AGNRs, remain for wider ribbons, where the effect of under-coordinated edge atoms reduced.
    The study of excitation due to radiation from an electromagnetic field or the application of a gate voltage is left for future research because our goal here is to present the potential of the proposed few-layer graphene heterostructures. 
	
	\section*{CRediT authorship contribution statement}
	
	M. B.: Conceptualization, Methodology, Investigation, Software, Writing – Original Draft.\\
	T. D. K.: Supervision, Funding Acquisition, Validation, Writing – Review and Editing.
	
	\section*{Acknowledgements}
	
	Part of the research was funded by the DFG via the CRC 1415 (project numbers 417590517).
	
	\section*{Declaration of competing interest }
	The authors declare that they have no known competing financial interests or personal relationships that could have appeared to influence the work reported in this paper.
	
	\section*{ Data availability }
    The data that support the findings of this study are available upon reasonable request from the authors.

	\bibliography{BIB.bib}                                             %
	
	\clearpage
	\appendix
	\renewcommand{\thefigure}{S\arabic{figure}}                    
	\setcounter{figure}{0}                                         

	\section*{Supplementary Material}
\begin{figure}[h]
		\centering
			\includegraphics[height=4.8cm,trim={0cm 0cm 0cm 0cm},clip]{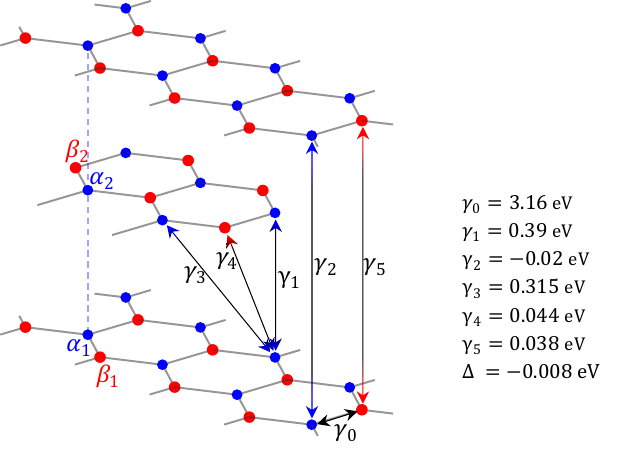}
		\caption{The SWMcC TB model is shown on 5A3-S atomic model. The $\alpha$ sublattices are vertically aligned, as shown by a blue dashed line. Also, the TB paramteres and their corresponding values are shown.}
		\label{figs1}
	\end{figure}

The atomic structure of a 5A3-S heterostructure and the corresponding SWMcC tight-binding model are shown in Fig.~\figref{figs1}{}. The two sublattices are depicted in blue and red, with the $\alpha$ (blue) sublattices vertically aligned.
	
\begin{figure}[h]
		\centering
		\begin{subfigure}[b]{0.54\linewidth}
			\caption{}
			\includegraphics[height=4.8cm,trim={5cm 7.7cm 5.9cm 7.7cm},clip]{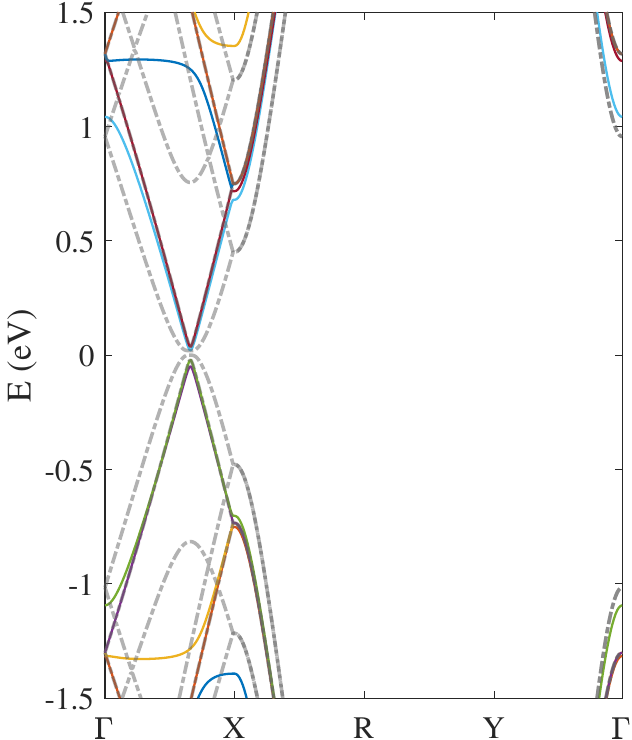}
			\label{s2a}
		\end{subfigure}
		\hfill
		\begin{subfigure}[b]{0.44\linewidth}
			\caption{}
			\includegraphics[height=4.8cm,trim={5cm 7.7cm 5.9cm 7.7cm},clip]{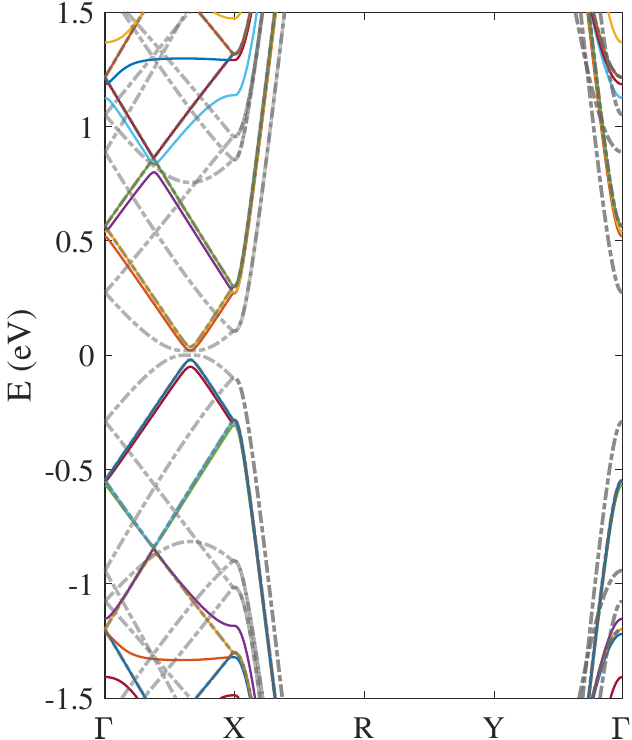}
			\label{s2b}
		\end{subfigure}
		\vfill
		\begin{subfigure}[t]{0.54\linewidth}
			\caption{}
			\includegraphics[height=4.8cm,trim={5cm 7.7cm 5.9cm 7.7cm},clip]{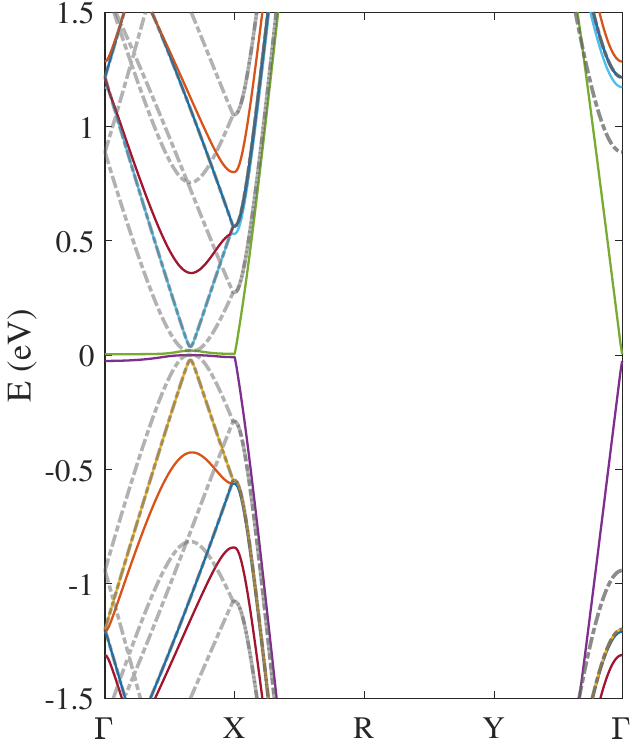}
			\label{s2c}
		\end{subfigure}
		\hfill
		\begin{subfigure}[t]{0.44\linewidth}
			\caption{}
			\includegraphics[height=4.8cm,trim={5cm 7.7cm 5.9cm 7.7cm},clip]{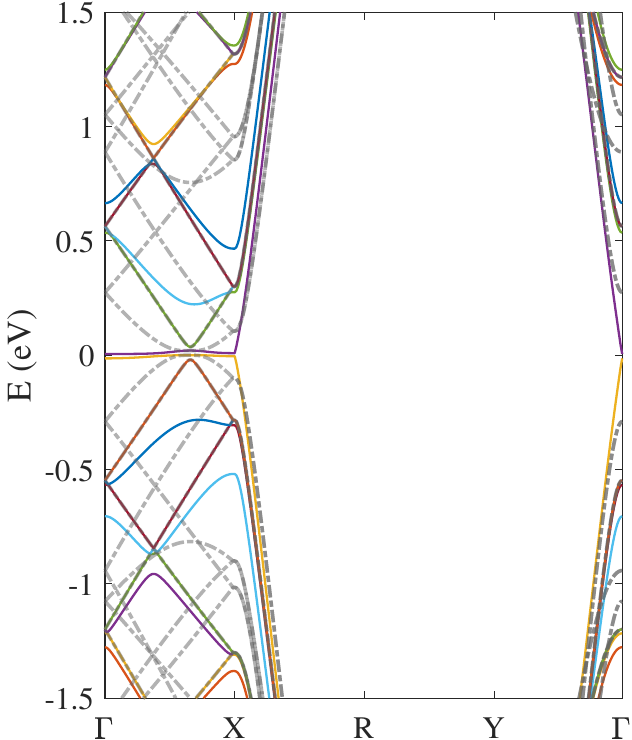}
			\label{s2d}
		\end{subfigure}
		\caption{The electronic band structure of (a) 3A5-S, (b) 3A17-S, (c) 5A5-S, and (d) 5A15-S. The band structure of the perfect cases are presented in dashed gray lines.}
		\label{figs2}
	\end{figure}

	The effect of spacing between AGNRs is weak. A gapped and gapless AGNR is studied with different spacings in Figure~\figref{figs2}{}.
    
	\begin{figure}[h]
	\centering
	\includegraphics[height=4.8cm,trim={5cm 7.7cm 5.9cm 7.7cm},clip]{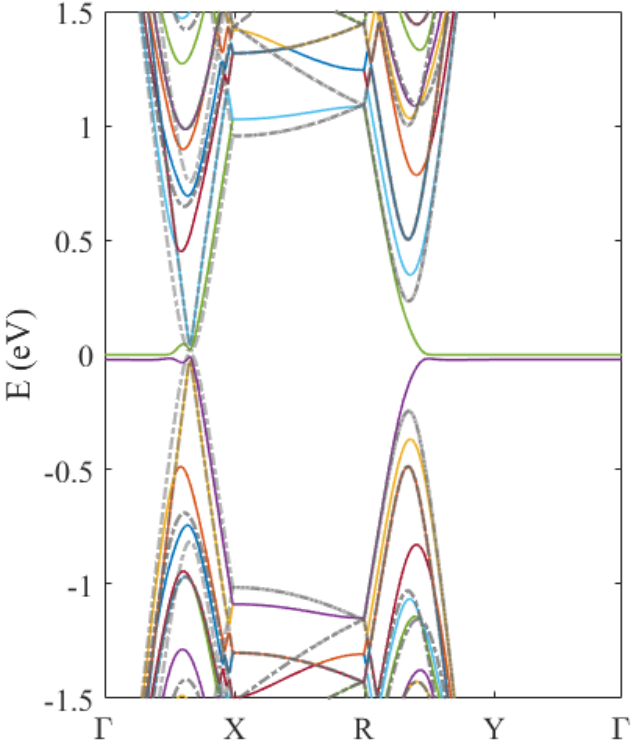}
	\caption{The band structure of 36Z4-S.}
	\label{figs3}
\end{figure}

	\begin{figure}[h]
		\centering
		\begin{minipage}{\figwidth} 
			\centering
			\begin{subfigure}[t]{0.52\figwidth}
				\caption{}
				\includegraphics[height=4.8cm,trim={5cm 7.7cm 5.9cm 7.7cm},clip]{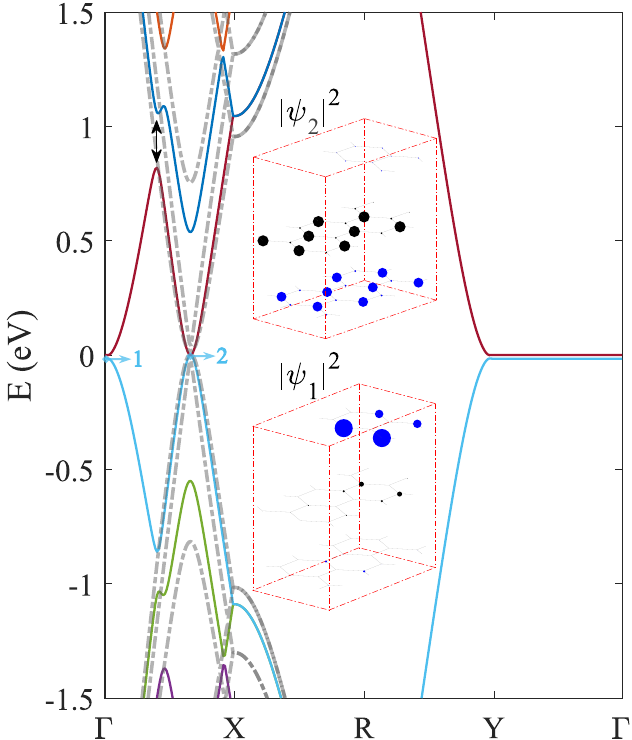}
				\label{s4a}
			\end{subfigure}
			\hfill
			\begin{subfigure}[t]{0.43\figwidth}
				\caption{}
				\includegraphics[height=4.8cm,trim={6.6cm 7.7cm 5.9cm 7.7cm},clip]{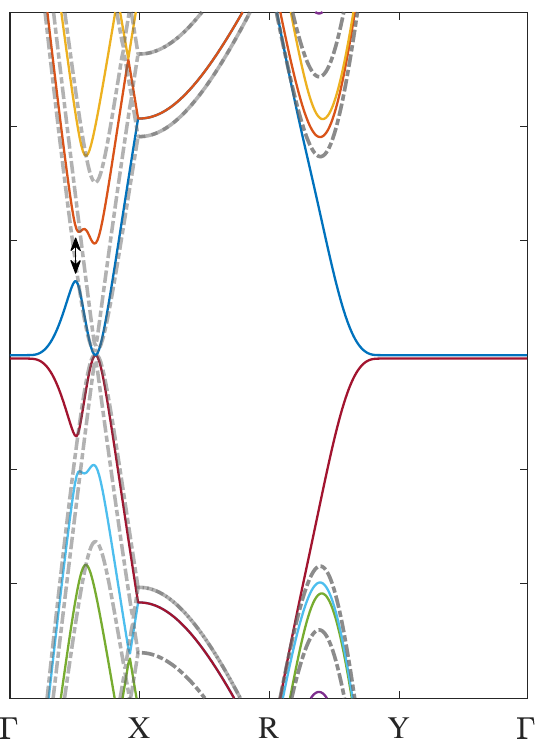}
				\label{s4b}
			\end{subfigure}
		\end{minipage}
		\caption{The band structure of (a) 4Z4-NS and (b) 10Z6-NS. The site-resolved probability density of the marked states are show in the insets. The gap between the 1st and 2nd conduction bands are indicated by double head black arrow.}
		\label{figs4}
	\end{figure}
	
	\FloatBarrier
	
    In the following, we applied a perpendicular electric field to 6A4 bilayer system. In the TB model, it is considered by changing the onsite energies of the AGNR layer to include a uniform electric field with a strength of 1 V/nm. Comparing \figref{figs6}{s6a} and \figref{}{s6b} shows opening a band gap upon applying the field.
	\begin{figure}[H]
		\centering
		\begin{minipage}{\figwidth} 
			\centering
			\begin{subfigure}[t]{0.52\figwidth}
				\caption{}
				\includegraphics[height=4.8cm,trim={4.8cm 8.5cm 5.6cm 8.6cm},clip]{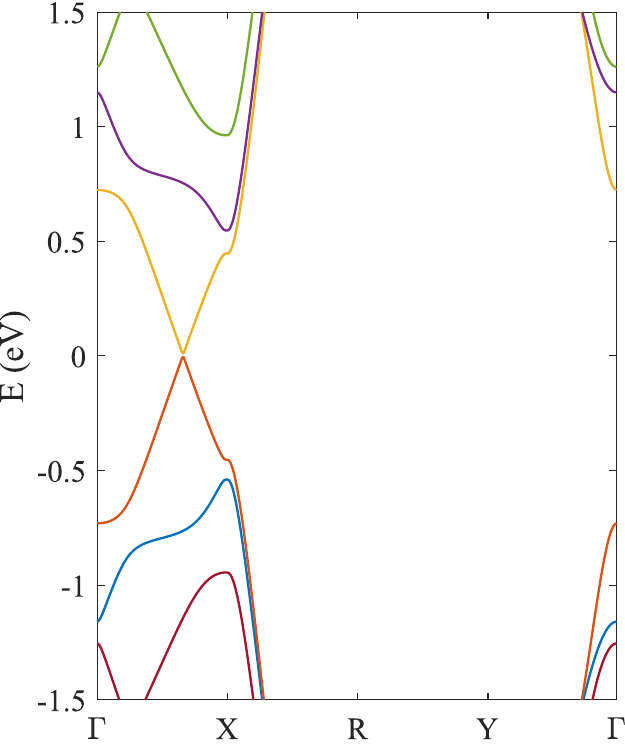}
				\label{s5a}
			\end{subfigure}
			\hfill 
			\begin{subfigure}[t]{0.43\figwidth}
				\caption{}
				\includegraphics[height=4.8cm,trim={6.3cm 8.5cm 5.6cm 8.6cm},clip]{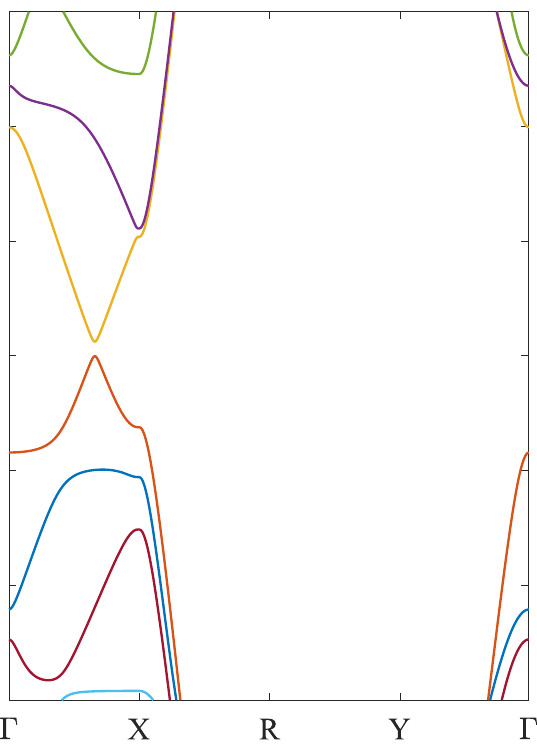}
				\label{s5b}
			\end{subfigure}
		\end{minipage}
		\caption{The band structure of 6A4 (a) without, and (b) with an applied electric field. }
		\label{figs5}
	\end{figure}

    	The band structure of 5A11-NS is observed to depend on the spacing between semimetal AGNRs. The band gap can become indirect, which originates from the hopping terms of non-adjacent layers (see Figure~\figref{figs5}{}). However, the conduction band minima and valence band maxima of the dispersive bands occur at nearly the same $\mathbf{k}$-point,indicating a negligible momentum difference . Such a small $\mathbf{k}$ difference does not make a meaningful change in physical properties; however, further studies may still be needed. 
	\begin{figure}[H]
		\centering
		\begin{minipage}{\figwidth} 
			\centering
			\begin{subfigure}[t]{0.52\figwidth}
				\caption{}
				\includegraphics[height=4.8cm,trim={5cm 7.7cm 5.9cm 7.7cm},clip]{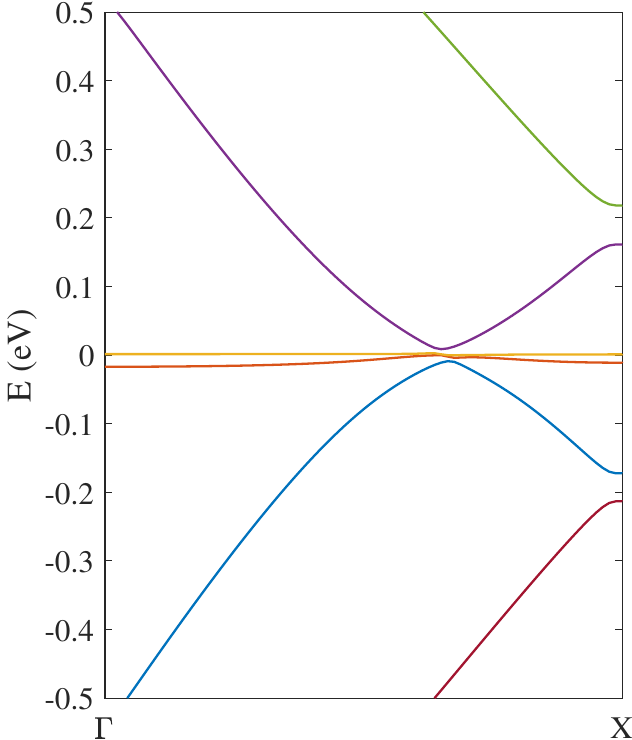}
				\label{s6a}
			\end{subfigure}
			\hfill 
			\begin{subfigure}[t]{0.43\figwidth}
				\caption{}
				\includegraphics[height=4.8cm,trim={6.6cm 7.7cm 5.9cm 7.7cm},clip]{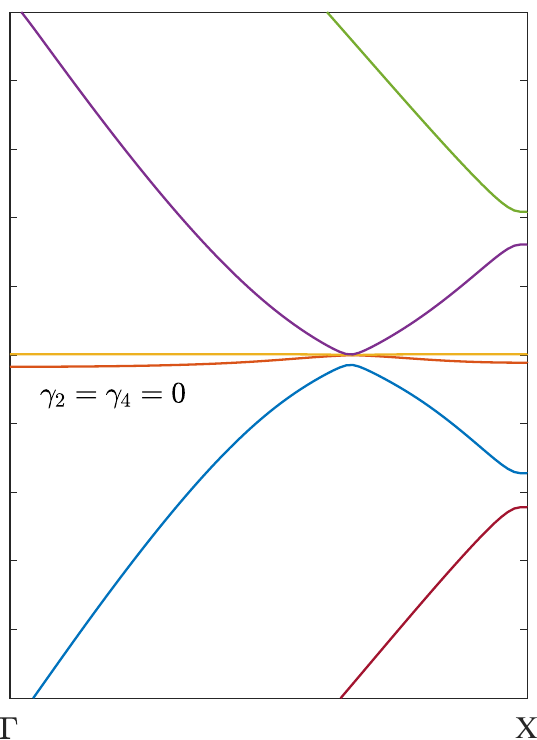}
				\label{s6b}
			\end{subfigure}
		\end{minipage}
		\caption{The band structure of 5A11-NS (a) with all hopping terms, and (b) by neglecting the coupling terms of non-adjacent layers.}
		\label{figs6}
        
	\end{figure}
\end{document}